# PARTICLE PHYSICS and the COSMIC MICROWAVE BACKGROUND

John E. Carlstrom, Thomas M. Crawford, and Lloyd Knox

**Temperature and polarization variations across the microwave sky include the fingerprints of quantum fluctuations in the early universe. They may soon reveal physics at unprecedented energy scales.**

Fifty years ago Bell Labs scientists Arno Penzias and Robert Wilson encountered a puzzling excess power coming from a horn reflector antenna they had planned to use for radio astronomical observations. After painstakingly eliminating all possible instrumental explanations, they finally concluded that they had detected a faint microwave signal coming from all directions in the sky.[1] That signal was quickly interpreted as coming from thermal radiation left over from a much hotter and earlier period in our universe's history, and the Big Bang was established as the dominant cosmological paradigm.[2] Cooled by the expansion of the universe to a temperature just below 3 K, so that its intensity peaks in the microwave region of the spectrum, the radiation detected by Penzias and Wilson is known today as the cosmic microwave background (CMB). The two scientists were awarded the 1978 Nobel Prize in Physics for their discovery.

The detection of the CMB and the consensus that the universe had a hot and dense early phase led to a fertile relationship between cosmology and particle physics. The hot early universe was a natural particle accelerator that could reach energies well beyond what laboratories on Earth will attain in the foreseeable future. Precise measurements of both the spectrum of the CMB and its tiny variations in brightness from one point to another on the sky reflect the influences of high-energy processes in the early cosmos.

For instance, the gravitational effects of neutrinos have been detected at high significance; such measurements imply that the sum of the neutrino masses is no more than a few tenths of an eV. The CMB data also show the influence of helium produced in the early universe and thus constrain the primordial helium fraction. Moreover, the data are nearly impossible to fit without dark energy and dark matter—two ingredients missing from the standard model of particle physics (see the article by Josh Frieman, PHYSICS TODAY, April 2014, page 28).

## Simple math, difficult observations

The constraining power of CMB observations follows in part from the high degree of isotropy in the CMB and the resulting precision with which theorists can offer predictions. The coupled, nonlinear equations for the evolving photon distribution are difficult to solve exactly. But because the temperature of the CMB is the same in every direction to about 1 part in 100 000, the linearized equations, which are relatively easy to solve, provide an excellent approximation. Of course, the small level of anisotropy has a downside: It makes measurement quite challenging. For nearly 30 years after the discovery of the CMB, observations yielded only increasingly stringent limits on its anisotropy. Then, in 1992, an instrument aboard the *Cosmic Background Explorer* (*COBE*) measured the anisotropy for the first time.[3]

Since the *COBE* era, observations of the CMB with dedicated, specialized telescopes have become orders of magnitude more sensitive, thanks to remarkable progress in the development of microwave detectors and measurement techniques. The 1990s saw experiments across a range of angular scales from the ground and from balloons; measurements of degree-scale anisotropy in particular helped establish the current cosmological picture of a geometrically flat universe dominated by dark

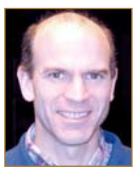
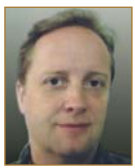
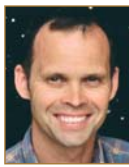

**John Carlstrom** is a professor of physics and of astronomy and astrophysics and deputy director of the Kavli Institute for Cosmological Physics (KICP) at the University of Chicago. **Tom Crawford** is a senior researcher at the KICP. **Lloyd Knox** is a professor of physics at the University of California, Davis.

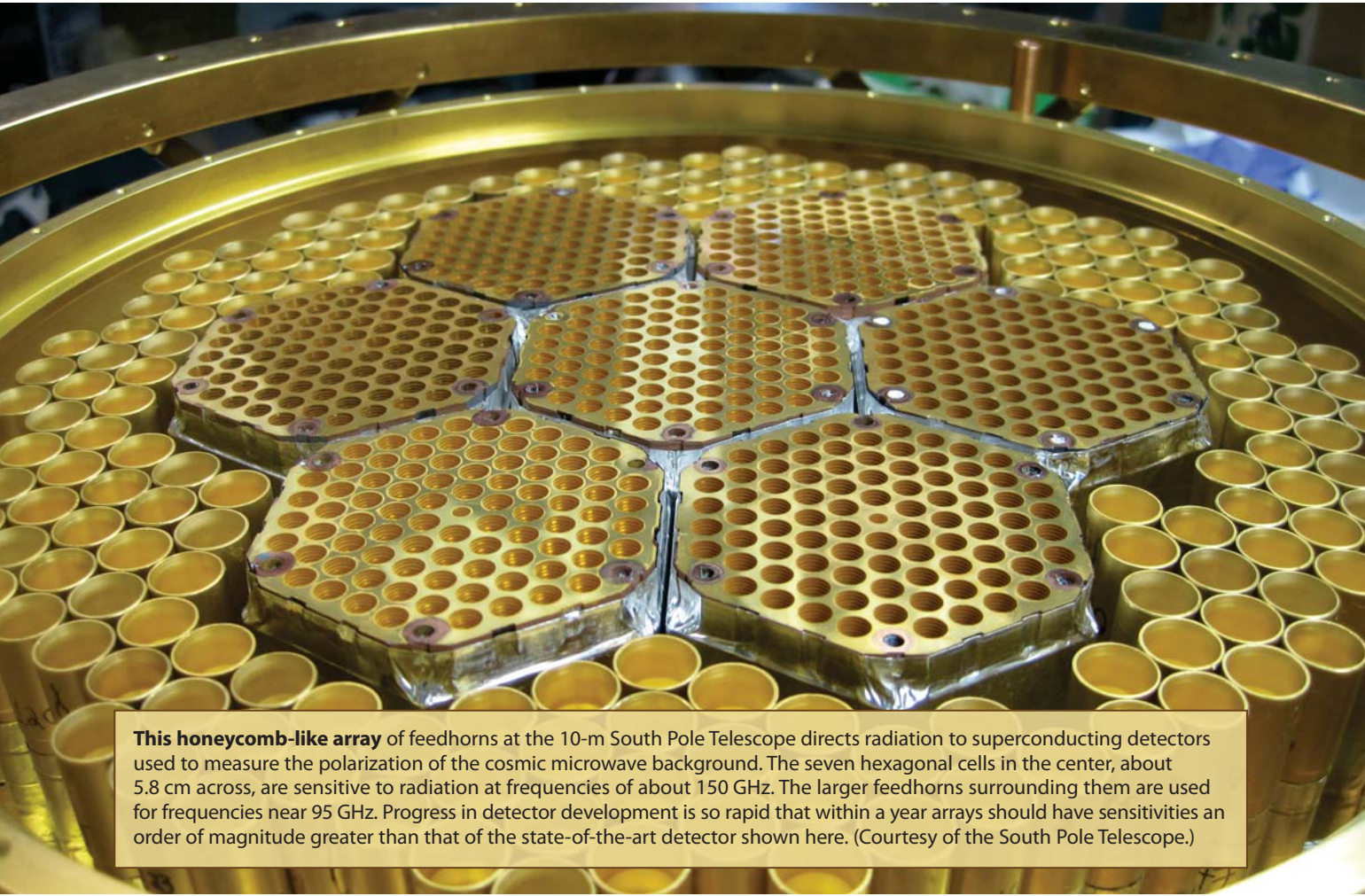

**This honeycomb-like array** of feedhorns at the 10-m South Pole Telescope directs radiation to superconducting detectors used to measure the polarization of the cosmic microwave background. The seven hexagonal cells in the center, about 5.8 cm across, are sensitive to radiation at frequencies of about 150 GHz. The larger feedhorns surrounding them are used for frequencies near 95 GHz. Progress in detector development is so rapid that within a year arrays should have sensitivities an order of magnitude greater than that of the state-of-the-art detector shown here. (Courtesy of the South Pole Telescope.)

energy and dark matter. The next decade saw the launch of two successor satellites to *COBE*, the *Wilkinson Microwave Anisotropy Probe* (*WMAP*) in 2001 and *Planck* in 2009.

These days the temperature of the CMB has been exquisitely mapped out by the satellite experiments and by large-aperture, ground-based telescopes such as the South Pole Telescope and the Atacama Cosmology Telescope. The new observational frontier is CMB polarization, first detected in 2002 by the DASI (Degree Angular Scale Interferometer) experiment at the South Pole[4] and soon thereafter by *WMAP*.[5] Since those initial measurements, the polarization signature of density fluctuations in the primordial plasma has been mapped with high signal to noise, and the effects of gravitational lensing on the polarization have been detected.

The anisotropy in the CMB is of crucial physical importance because those tiny temperature fluctuations reflect the small density inhomogeneities that, under the influence of gravity, grew over time to become all the structures we see in the universe. Without an anisotropic CMB, we would not exist. But what was the origin of the small fluctuations? A compelling answer comes from the next great synergy between the CMB and particle physics: the theory of inflation.

### In and out of causal contact

Once a highly speculative idea, inflation is now a part of the standard cosmology. Although deep questions about inflation remain, the predictions of the simplest versions of the theory have been so successful that most cosmologists accept that some form of inflation truly did occur in the very early universe. The 2014 Kavli Prize in Astrophysics was recently awarded to Alan Guth, Andrei Linde, and Alexei Starobinsky for their pioneering contributions to the theory of cosmic inflation. Several others made critical contributions as well. For a first-hand account of the discovery and early development of the theory see reference 6.

Inflation is, by definition, a period of accelerating expansion. As explained in figure 1, an accelerating cosmos has a causal structure very different from that of a decelerating cosmos. In a decelerating universe, a pair of separated points evolves from being causally disconnected—in which case the particles, unable to influence each other, are said to be superhorizon, or separated by a horizon—to being causally connected, or subhorizon. In an accelerating universe, the opposite occurs. In the inflationary scenario, the universe undergoes an accelerating stage, which is followed by a long period of deceleration.

In view of the early period of accelerating expansion, two separated regions in the universe that are now causally disconnected could have been able to interact with each other during the inflationary epoch. Causally connected perturbations in those two regions—for example, an underdensity in one and an overdensity in the other—could thus have been created at very early times. As we will see, quantum mechanics provides a mechanism for generating such perturbations. Quantum mechanical fluctuations initially created with subnuclear wavelengths are stretched by the cosmic expansion to millimeter length scales within a tiny fraction of a second; at present they are astrophysically large. Thus observations of cosmic structure give physicists an opportunity to probe physics on extremely small length scales.

**Figure 1. In an expanding universe**, the distance between two separated points increases over time, simply due to the expansion of the space between them. The two panels here show the spacetime trajectories of two points, A and B. For the decelerating expansion illustrated in the top panel, the separation rate is greater in the past and even exceeds the speed of light at sufficiently early time. Thus A and B go from being out of causal contact—unable to influence each other—to being in causal contact. In an accelerating cosmos, the separation rate is smaller in the past; the two points go from being in causal contact to being out of causal contact. In the inflationary universe scenario, an early epoch of acceleration—the inflationary era—smoothly maps onto a long period of deceleration. Thus two points can go from being in causal contact to out of causal contact and, much later, back into causal contact. (Courtesy of Marius Millea.)

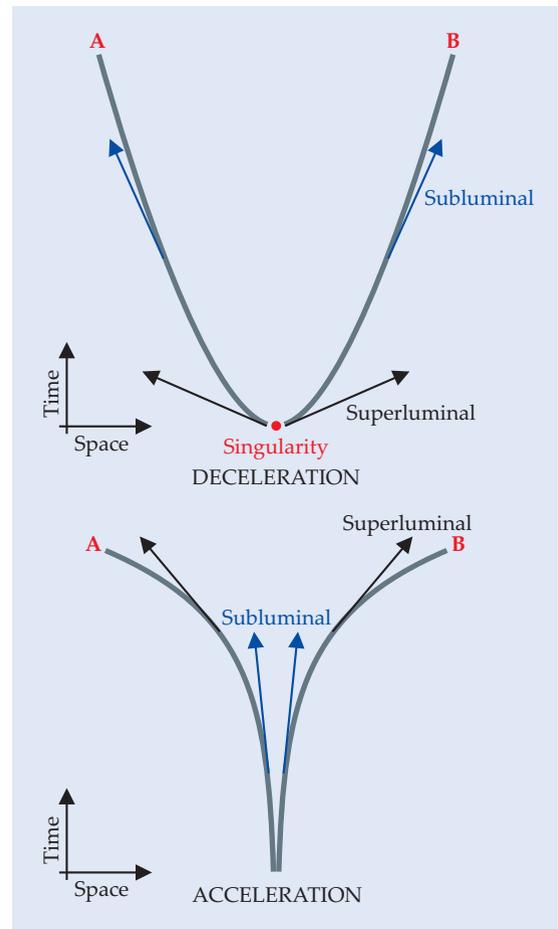

### A recipe for accelerated expansion

Accelerating expansion requires the universe to have an energy density that dilutes relatively slowly with expansion. In inflationary models, such an energy density is usually obtained via the introduction of a new field $\phi$, called the inflaton field. Just as the dynamics of ordinary particles are governed by a potential energy, so the dynamics of the inflaton field are governed by what particle physicists call a potential $V(\phi)$. (It's really a potential energy density.)

The potential $V(\phi)$ is only one of two contributions to the energy density of the inflaton field. The second contribution, a four-dimensional kinetic energy, arises from temporal and spatial derivatives. The total energy density of $\phi$ is thus

$$\rho = V(\phi) + \frac{1}{2}\left(\frac{d\phi}{dt}\right)^2 + \frac{c^2}{2}\nabla\phi \cdot \nabla\phi.$$

A generic inflaton field will not lead to inflation. But if the gradients in the field are small enough over a large enough patch of space and if the value of $\phi$ is sufficiently far away from the value that minimizes the potential, then $\phi$ will rapidly evolve to satisfy something called the slow-roll condition, $\frac{1}{2}(d\phi/dt)^2 \ll V(\phi)$. When both the spatial and temporal derivatives of the inflaton field are small, $V(\phi)$ is nearly constant in time and makes the dominant contribution to the energy density. Under such conditions, the patch inflates.

That the inflation results in a patch of essentially constant energy density follows from the general relativistic result that the Hubble parameter $H \equiv a^{-1} da/dt$ is proportional to $\rho^{1/2}$; here $a$ is the so-called scale factor that describes the size of the universe. Since $H$ is nearly constant, $da/dt$ increases as $a$ does—that is, the cosmic expansion accelerates. In the limit where the energy density, and thus $H$, is truly constant in time, the scale factor grows exponentially, proportional to $e^{Ht}$. In that limit, the horizon length is exactly $c/H$, with $c$ being the speed of light. In other words, points separated by more than $c/H$ are out of causal contact.

As inflation proceeds, the gradients of the inflaton field are stretched out by the expansion and the field becomes very smooth. If the inflationary epoch lasts long enough for the scale factor to increase by about $e^{60}$, then any initial irregularities will be stretched out to length scales that today are unobservably large; the result is a smooth observable universe with negligible spatial curvature. In one simple model of inflation, the potential satisfies $V(\phi) \propto \frac{1}{2} m^2 \phi^2$, where $m$ is the mass of the inflaton. In that realization, the entire observable universe once existed in a patch with a diameter of less than $10^{-29}$ m, or $10^{-14}$ of a proton radius. The total mass energy in that patch was about $10^4$ J, the caloric content of two diet Cokes. Today the observable universe includes regular Cokes and about $10^{70}$ J of mass energy from other sources, a result compatible with the perhaps surprising fact that energy is not conserved in general relativity.

### We are quantum fluctuations

Quantum mechanics limits how smooth the inflaton field can be; fluctuations in the field necessarily persist at a level dictated by the uncertainty principle. But as figure 2 shows, those fluctuations, too, will be stretched to astrophysically large length scales by cosmic expansion; thus quantum fluctuations provide the initial seeds of all structure in the universe. Carl Sagan popularized the notion that we are star stuff. The idea that "we are quantum fluctuations" deserves to be in the zeitgeist as well.

As $\phi$ rolls toward the potential minimum, $V(\phi)$ eventually becomes smaller than $\frac{1}{2}(d\phi/dt)^2$; the slow-roll condition is no longer met, and inflation ends. Decays of the inflaton to other particles—irrelevant during inflation because the decay products were quickly diluted by expansion—then become important. The remaining energy in the $\phi$ field converts to a thermal bath of the particles of the standard model, and perhaps other particles as well.

The small but nonzero spatial fluctuations in $\phi$, stretched from quantum to astrophysical scales by cosmic expansion, cause inflation to end at different times in different locations. In those regions where inflation ends relatively early, the mass density is lower due to the extra expansion that the region has undergone since the end of inflation. Thus the slightly different expansion histories of different locations result in density differences; those small density perturbations eventually grow under the influence of gravity to create all the structures we observe in the universe today.

Figure 3, of the *Hubble Space Telescope*'s Ultra Deep Field, is an iconic representation of that cosmic structure. The galaxies seen there—and, at smaller scales, stars, planets, and people—demonstrate that today the universe is anything but smooth. Images such as the *Hubble* Ultra Deep Field contain an immense quantity of information. Unfortunately, relating that information to theories of the early universe is difficult because the evolution of cosmic structure is complicated and nonlinear.

### The CMB power spectrum

Most cosmologists take inflation seriously because, as we will discuss in the following section, the theory has offered numerous successful predictions. Those predictions have been tested primarily via observations of the CMB, whose temperature fluctuations are shown in figure 4. That state-of-the art image was made with data from the *Planck* satellite.

The CMB photons last interacted with matter when the universe was just a few hundred thousand years old. Up to that time, the universe was dense and hot enough that no neutral atoms could survive. Instead, the cosmos was filled with a foggy plasma of photons, electrons, and protons strongly coupled through photon–electron scattering and Coulomb interactions. The tight coupling also meant that matter beginning to compress under the influence of gravity would bounce back due to the pressure support of the photons, a process that resulted in acoustic oscillations in the plasma.

When the universe cooled sufficiently, the protons and electrons combined into hydrogen atoms. At that "decoupling" time, the universe became transparent. The CMB photons we see today thus provide us with an image of a spherical shell, called the last scattering surface, sufficiently distant from us that photons headed in our direction from it are just arriving here today, 13.8 billion years later.

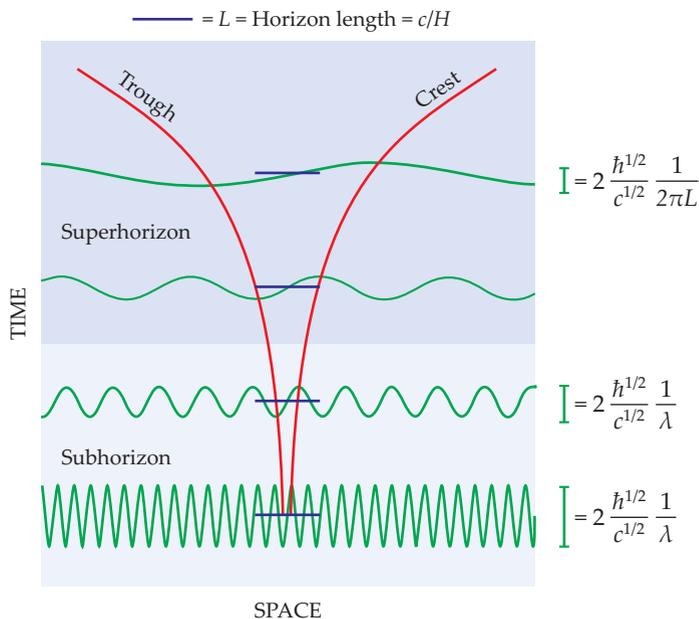

**Figure 2. Fluctuations in the value of the inflaton** field, which is responsible for the accelerating expansion of the cosmos, evolve differently, depending on whether their wavelength $\lambda$ is less than or greater than the horizon length $L = c/H$. (Here $c$ is the speed of light and $H$ is the Hubble parameter.) When $\lambda \ll L$, the expansion (red) smooths out the field, but the quantum uncertainty principle limits how smooth the field can be. As a result, the amplitude of the fluctuation is inversely proportional to $\lambda$ and thus decreases as the universe expands. (The influence of the uncertainty principle is reflected by the appearance of Planck's constant $\hbar$ in the expression for the amplitude.) As $\lambda$ becomes larger than the horizon, the crest and trough of the wave cease to be in causal contact, so the amplitude stops evolving. For superhorizon evolution, its asymptotic value corresponds to replacing the wavelength in the subhorizon case with $2\pi L$. Eventually, cosmic expansion stretches the fluctuations to astrophysically large length scales.

The most informative statistic of the CMB is its angular power spectrum, shown in figure 5. Roughly speaking, the power spectrum shows the anisotropy as a function of angular scale. To actually obtain the fluctuation power displayed in the figure, one decomposes the CMB map of the sky into spherical harmonic modes $Y_{\ell m}$ with complex coefficients $a_{\ell m}$. The fluctuation power $C_\ell$ for a specified value of $\ell$ is then given by

$$C_\ell = \frac{1}{2\ell + 1} \sum_{m=-\ell}^{\ell} |a_{\ell m}|^2.$$

The spherical harmonic $Y_{\ell m}$ executes $\ell$ cycles in 360°, so hot and cold spots separated by a degree correspond to modes with $\ell \approx 180$.

### Predictions of inflation

Inflation is not merely a theory that was constructed to be consistent with preexisting facts. It has made several falsifiable predictions. In this section we describe some of the most important ones.

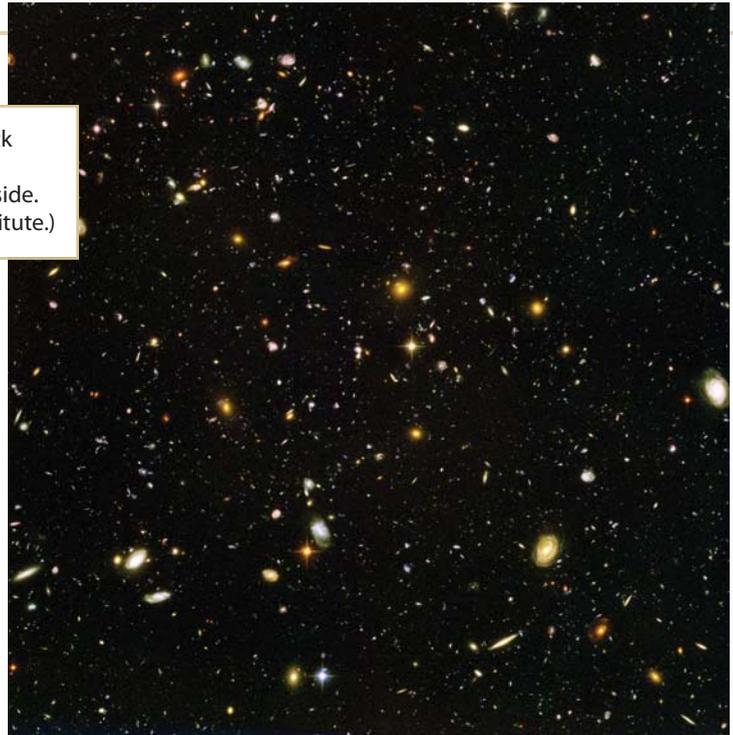

Figure 3. The *Hubble* **Ultra Deep Field** image looks back as much as 13 billion years. This image, from the *Hubble Space Telescope*, is approximately 3 arcminutes on each side. (Courtesy of NASA and the Space Telescope Science Institute.)

▶ **Small spatial curvature.** Any curvature of space present as inflation begins is rapidly expanded away, just as the curvature of a balloon decreases as the balloon is inflated. Thus the radius of curvature, the distance over which effects of nonzero curvature become significant, should be very large. Current CMB measurements show that the curvature radius is at least four times the radius of the observable universe.

▶ **A nearly scale-invariant spectrum of density perturbations.** To the extent that the energy density during inflation remains constant, an inflating patch of the cosmos looks the same at all times and the horizon length $L = c/H$ is fixed. As illustrated in figure 2, all mode amplitudes take on nearly the same value once the wavelength has stretched out enough to be significantly larger than $L$. The primary difference between modes that became superhorizon earlier is that they have undergone more expansion since horizon crossing and thus have longer wavelengths.

Inflation cannot be completely time-translation invariant because it has to end. If the terminating transition is smooth, we should see evidence that the average value of the mode amplitudes should vary slightly with wavelength. Strong evidence for such a departure from scale invariance has indeed been found through analysis of the CMB angular power spectrum.

▶ **Acoustic peaks in the CMB angular power spectrum.** According to inflation theory, acoustic oscillations in the primordial plasma will result in a series of peaks in the CMB angular power spectrum. As figure 5 shows, those peaks have now been well measured.

The period of superhorizon evolution in the inflationary scenario provides special initial conditions for the acoustic oscillations—namely, all modes with the same wavelength begin to oscillate at the same time with zero initial momentum. As a result of those special conditions, the primordial plasma includes a set of standing waves for which all modes of a given wavelength have the same initial phase. Because modes of a given wavelength all oscillate at a specific rate governed by the sound speed in the plasma, they remain phase synchro-nized for all time. The result is a series of acoustic peaks in the CMB power spectrum, in which the $n$th peak corresponds to modes that arrived at their $n$th extremum at the decoupling time.

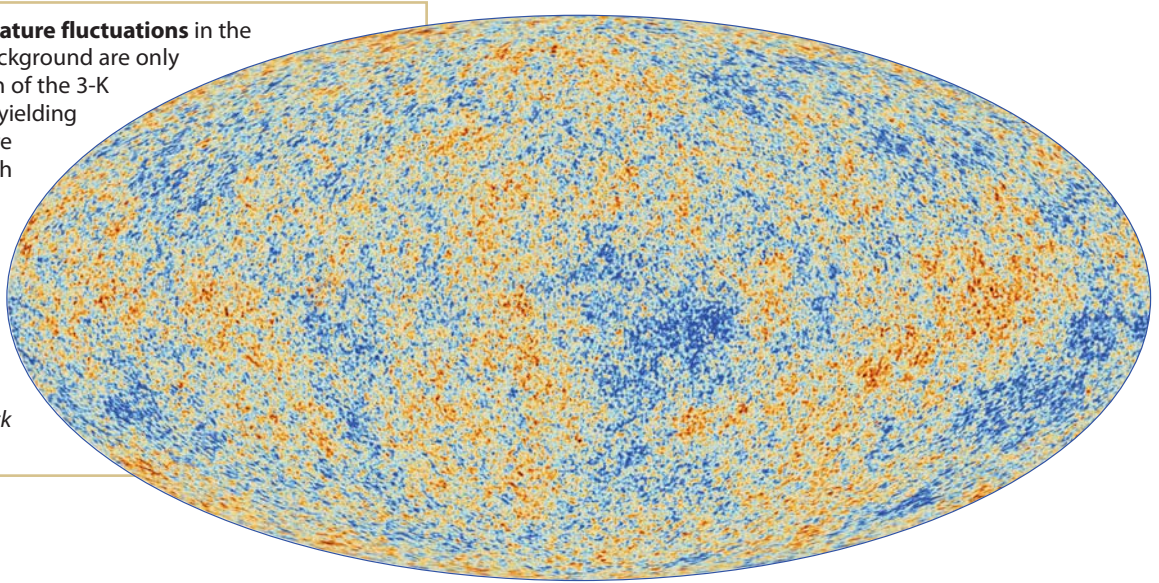

Figure 4. **The temperature fluctuations** in the cosmic microwave background are only a few parts per million of the 3-K mean value. The data yielding this image of the entire sky were obtained with instruments aboard the *Planck* satellite. A single pixel on this map covers more of the sky than the *Hubble* Ultra Deep Field shown in figure 3. (Courtesy of the *Planck* collaboration.)

▶ **Gaussian perturbations.** The expectation that energy density perturbations are Gaussian follows from the wavefunction of a harmonic oscillator in its ground state being Gaussian. To the extent that density fluctuations $\delta\rho$ depend linearly on $\delta\phi$, the density perturbations will be Gaussian as well. For an essentially constant inflaton field, the energy density is determined solely by the potential $V(\phi)$. The Taylor-series expansion $\delta\rho = (dV/d\phi)\,\delta\phi + \frac{1}{2}\,(d^2V/d\phi)^2\,\delta\phi^2$ indicates that a nonzero second derivative causes a non-Gaussian $\delta\rho$. However, a large second-derivative term for the potential would also ruin the slow-roll behavior of the inflaton field; such considerations greatly restrict the amount of non-Gaussianity allowed in the simplest models. The *Planck* data have dramatically improved the sensitivity of searches for non-Gaussianity. Quadratic corrections are limited to be less than about $10^{-4}$ times the linear, Gaussian term.[7]

## The hunt for gravitational waves

One key prediction of inflation that remains unconfirmed is the existence of a nearly scale-invariant spectrum of gravitational waves—degrees of freedom in the spacetime metric that can be excited without any corresponding excitation of matter fields. Just as with fluctuations of the inflaton field, they obey an uncertainty principle and, in the course of superluminal expansion, have their amplitude set to a value proportional to the Hubble parameter $H$ during inflation. Detecting the influence of that gravitational-wave background on the CMB would allow cosmologists to infer $H$ and hence the energy scale of the inflationary potential; observations of density perturbations, by contrast, provide a relatively indirect look at the inflationary era.

Attempts to measure the effect of gravitational waves on the CMB's temperature fluctuations have produced upper limits only. However, as the electrons and protons in the primordial plasma combine into hydrogen, any gravitational waves that are around will produce unique signatures in the polarization of the CMB—divergence-free patterns, called *B* modes, that, to linear order, cannot be created by density perturbations.[8] To better constrain the inflationary epoch, scientists will need to measure the CMB polarization and tease out the signature of gravitational waves.

Several ground- and balloon-based experiments currently under way are attempting to do just that. The instruments the researchers are using are so sensitive that the studies are limited by systematic effects. For example, in March of last year, the BICEP2 collaboration announced it had detected *B* modes whose power spectrum had an angular dependence consistent with inflation.[9] Subsequent data from the *Planck* collaboration[10] and, most recently, a collaborative cross-correlation of BICEP2 and *Planck* data sets[11] have demonstrated that the signal originally reported by BICEP2 is consistent with having arisen entirely from dust emission in our own galaxy. (For more, see PHYSICS TODAY, May 2014, page 11, and the Commentary by Mario Livio and Marc Kamionkowski, December 2014, page 8.)

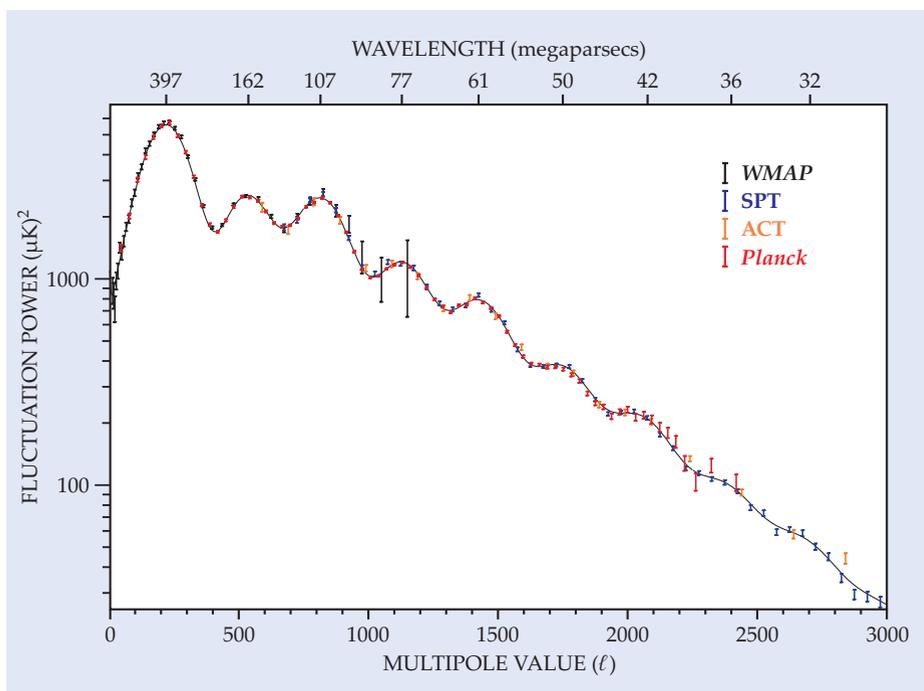

**Figure 5. The angular power spectrum** of the cosmic microwave background (CMB) displays a series of acoustic peaks, as predicted by inflation theory. A peak at a multipole value of $\ell$ means that the fluctuations in the CMB include a significant component of hot spots and cold spots separated by $180/\ell$ degrees. The first peak corresponds to acoustic oscillations (defined in the text, which also gives a precise definition of the fluctuation power) that reached their first extremum about 370 000 years after the Big Bang, when the universe decoupled, or became transparent to photons. As indicated on the upper axis, those modes now have a wavelength of about 400 megaparsecs ($1.3 \times 10^9$ light-years). Modes with a current wavelength of 162 Mpc oscillated faster and achieved their second extremum at decoupling. Between those two wavelengths are modes with $\ell \approx 400$ that hit a null in their oscillations at decoupling; those modes are responsible for the trough at 213 Mpc. The data shown here were obtained by the *Wilkinson Microwave Anisotropy Probe* (*WMAP*), the South Pole Telescope (SPT), the Atacama Cosmology Telescope (ACT), and the *Planck* satellite. The curve is the prediction of a representative inflationary model. (Courtesy of Zhen Hou.)

The excitement surrounding searches for inflationary gravitational waves has not abated. After all, a detection of gravitational waves from inflation would open a window to physics at energy scales a trillion times larger than those accessible at the Large Hadron Collider and provide the first observational evidence that the gravitational field is a quantum field. Even as current experiments run and others are being built, physicists in the US are planning an ambitious next-generation experiment that will use telescopes at the South Pole, the high Atacama plateau in Chile, and possibly Northern Hemisphere sites to image the polarization over most of the sky with unprecedented sensitivity and in multiple frequency channels. In addition, the international CMB community is working on satellite mission concepts. All of those efforts are aiming for a sensitivity to the inflationary $B$-mode signals at a level of 1% of the current upper limits. Success in those endeavors would yield additional science dividends, including a determination of the absolute masses of neutrinos and an exquisitely sensitive profile of the contents of the primordial plasma.[12]

Thirty years ago inflation was a highly speculative idea about the origin of the Big Bang, born from the application of modern ideas about particles and fields to questions about the early evolution of the universe. With its empirical successes, inflation is by consensus the best paradigm—notwithstanding some notable dissenting views—for the mechanism that generated the primordial density fluctuations that led to all structure in the universe. Its success has motivated physicists to search for the siblings of those fluctuations, the gravitational waves, via their signature in the polarization of the CMB. If discovered, that gravitational imprint would open up an observational window onto quantum gravitational effects, extremely early times, and extremely high energies.

*We thank Brent Follin, Daniel Green, Zhen Hou, Ryan Keisler, Marius Millea, Rajiv Singh, and Gergely Zimanyi for their comments.*

### References

1. A. A. Penzias, R. W. Wilson, *Astrophys. J.* **142**, 419 (1965).
2. R. H. Dicke et al., *Astrophys J.* **142**, 414 (1965).
3. G. F. Smoot et al., *Astrophys. J. Lett.* **396**, 1 (1992).
4. J. M. Kovac et al., *Nature* **420**, 772 (2002).
5. A. Kogut et al., *Astrophys. J. Suppl. Ser.* **148**, 161 (2003).
6. A. H. Guth, *The Inflationary Universe: The Quest for a New Theory of Cosmic Origins*, Addison-Wesley (1997).
7. For an overview of the *Planck* mission and the cosmology results from its February 2015 data release, see R. Adam et al. (*Planck* collaboration), http://xxx.lanl.gov/abs/1502.01582.
8. U. Seljak, M. Zaldarriaga, *Phys. Rev. Lett.* **78**, 2054 (1997); M. Kamionkowski, A. Kosowsky, A. Stebbins, *Phys. Rev. Lett.* **78**, 2058 (1997).
9. P. A. R. Ade et al. (BICEP2 collaboration), *Phys. Rev. Lett.* **112**, 241101 (2014).
10. R. Adam et al. (*Planck* collaboration), http://arxiv.org/abs/1409.5738.
11. P. A. R. Ade et al. (BICEP2–Keck and *Planck* collaborations), *Phys. Rev. Lett.* (in press).
12. K. N. Abazajian et al., *Astropart. Phys.* **63**, 55 (2015); K. N. Abazajian et al., *Astropart. Phys.* **63**, 66 (2015). ∎